# A multi-view approach for Mandarin non-native mispronunciation verification

*Zhenyu Wang*[1,2], *John H. L. Hansen*[2], *Yanlu Xie*[1]

Beijing Language and Culture University, Beijing, China[1],
Center for Robust Speech Systems (CRSS), University of Texas at Dallas, U. S. A[2]

`{zhenyu.wang, john.hansen}@utdallas.edu, xieyanlu@blcu.edu.cn`

## Abstract

Traditionally, the performance of non-native mispronunciation verification systems relied on effective phone-level labelling of non-native corpora. In this study, a multi-view approach is proposed to incorporate discriminative feature representations which requires less annotation for non-native mispronunciation verification of Mandarin. Here, models are jointly learned to embed acoustic sequence and multi-source information for speech attributes and bottleneck features. Bidirectional LSTM embedding models with contrastive losses are used to map acoustic sequences and multi-source information into fixed-dimensional embeddings. The distance between acoustic embeddings is taken as the similarity between phones. Accordingly, examples of mispronounced phones are expected to have a small similarity score with their canonical pronunciations. The approach shows improvement over GOP-based approach by +11.23% and single-view approach by +1.47% in diagnostic accuracy for a mispronunciation verification task.

*Index Terms*—phone embedding, Siamese networks, mispronunciation verification, computer-assisted language learning, neural networks

## 1. Introduction

In previous studies, ASR have been a necessary component of computer-assisted language learning (CALL) systems to automatically assess proficiency of non-native speakers. Witt & Young introduced a likelihood-based "Goodness of Pronunciation" (GOP) measure [1] considering the likelihood of both canonical phone and a set of competing phones by human judges. Frame-level Log-posterior probabilities produced by an ASR component were averaged in a segment to represent confidence scores of each phone, which was a variant of GOP scores proposed by [2].

It is noted that this CALL systems' performances [1] is deeply dependent on the quality of non-native corpora labeling at the phone level to train models for phone-level confidence scores. Meanwhile, non-native learners' pronunciation is expected to be more native with reduced constraints from their primary language (L1) [3]. Non-native learners' mispronunciations always contain "distortion errors" such that intermediate states cannot be straightly deemed as phonemic substitution [4,5], which causes the complexity for annotators to provide ground-truth labels in some cases.

Therefore, weak supervision was applied to obtain embeddings in the form of discriminative feature representations to lift restriction of sparse mispronunciation annotation. This weak supervision approach can also be put into use in low-resource situations [6,7]. Bengio & Heigold employed a convolution neural network to embed word-level acoustic information for rescoring a speech recognizer' outputs based on a loss with a combination of ranking criteria and classification [8]. Chen et al. employed LSTM networks to embed acoustic words for a keyword spotting task [9], using a classification loss. Audhkhasi et al. trained auto-encoders for acoustic and written words respectively, and developed a comparison model on the two perspectives, which was also used for a key word search task [10]. Studies on acoustic word embeddings have focused on a number of embedding models, training approaches and tasks [11,12,13,14,15].

Systems noted above only adopt acoustic features as their inputs, however, there are other features and patterns which can be used to describe acoustic traits at the phone-level. In this study, Pronunciation is taken as a term of a sequence of phones in an utterance without construction information including the choice of syllables and prosody. Our approach is introduced for learning acoustic phone embeddings in a multi-view setting, which is similar to [16] but applied instead for a word discrimination task. Pronunciation dissimilarity between native speakers and non-native speakers can be taken as an ideal predictor of pronunciation goodness. By applying a multi-view scenario, acoustic sequences and multi-source information related to pronunciation can be jointly projected into a high-level representation space, where we can obtain acoustic embeddings of phones and use the distance between acoustic embeddings to represent pronunciation similarity. Given the mismatch between native and non-native speech, it is a very strong assumption that all phonetic embedding dissimilarity measures are exclusively interpreted as mispronunciations, but not phone-level recognition errors due to generic acoustic modelling confusion resulting from the mismatch. So it is suggested to replace traditional likelihood-based measures with embedding-based measures to improve mispronunciation verification. Multi-source information includes knowledge-based speech attributes and data-based bottleneck features, which help acoustic embedding models to produce a more discriminative representation of acoustic sequences. The learned acoustic embeddings are also expected to better represent the way phones sound from a perceptual view point. Several contrastive losses corresponding to different objectives for embedding learning are employed to optimize the distance between embeddings.

## 2. Overview of verification framework

### 2.1. Goodness of pronunciation score calculation

The GOP calculation form here is adopted from [2]. Eq. (1) was used to calculate the log posteriors of phone *p*:

$$log\, P(p|o; t_s, t_e) = \frac{1}{t_e - t_s} \sum_{t=t_s}^{t_e} log \sum_{s \in p} P(s|o_t) \quad (1)$$

where $o_t$ is the acoustic feature at frame $t$; the start time and the end time of phone $p$ are $t_s$ and $t_e$ respectively, derived from forced-alignment. $P(s|o_t)$ is the posterior at the frame level; $s$ is the context-dependent label; set $\{s \in p\}$ is a pool of context-dependent phones with phone $p$ in the central position. The GOP score of phone $p$ can be calculated by Eq. (2):

$$GOP(p) = \log \frac{P(p|o; t_s, t_e)}{max_{\{q \in Q\}} P(q|o; t_s, t_e)}, \quad (2)$$

Where p is the canonical phone, $q$ is the competing phone, and $Q$ is a pool of possible phones. A threshold is needed to verify whether the current phone is a mispronunciation.

### 2.2. Single-view approach with phone embedding and Siamese networks

The traditional GOP models require labeled non-native speech data added into the training data for adapting models to evaluate non-native learners' pronunciation, and very high diagnostic accuracy is needed to advance the solution. It is also of interest to apply a weak supervision approach based on learning from pairs of acoustic examples, using a contrastive loss. Pair-wise labels are a type of side information indicating whether the paired data is the same or not, which is easy to obtain in low-resource and data-sparse situations. Unidentified matching pairs can typically be found by an unsupervised term discovery system based on previous studies [17,18]. Such pair-wise supervision methods have introduced the Siamese network to adapt to a discrimination objective. The Siamese network as suggested in [19], which has been used for various domains including vision applications [20] and semantic word embedding [21,22,23]. The network in this module consists of three identical neural networks with tied parameters, taking three chunks of acoustic features as input and projecting input into embeddings formed by the last fully-connected layers. The training objective is to optimize the distance between embeddings with a contrastive loss such that the embeddings corresponding to the same phones are close, and embeddings of different phones are far from each other.

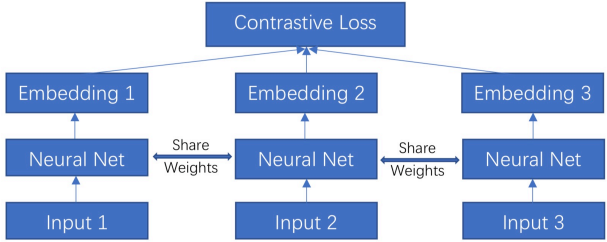

Figure. 1: *Siamese networks structure.*

Fig. 1 illustrates the structure of the Siamese network with three inputs, where input1 and input2 are both acoustic feature matrix from the same type of phones, while input3 is a different feature matrix. A contrastive loss is listed below, similar to that of [24], employed to serve the objective of projecting the acoustic features into embeddings in the high-level representation space.

$$Loss_{contrast} = \max\{0, m + d_{cos}(x_1, x_2) - d_{cos}(x_1, x_3)\}$$

$$\text{with } d_{cos}(x_1, x_2) = 1 - \left(\frac{<x_1, x_2>}{||x_1|| \cdot ||x_2||}\right)^2 \quad (3)$$

This loss is based on cosine distance, which aims to optimize the angle between embeddings corresponding to the same phone which ideally could be zero, and the angle for distinct phones would be orthogonal.

### 2.3. Multi-view approach with multi-source information and embedding model

The single-view approach take no advantage of multi-source information contained in bottleneck features and speech attribute patterns corresponding to phone-level segments. Here, we use a multi-view approach to learn embeddings from acoustic feature and multi-source information. The multi-view training framework is shown in Fig. 2.

Bottleneck feature is a data-driven feature reflecting pronunciation information which contributes to phone discrimination. Bottleneck features are the outputs from internal layers in a multi-layer perceptron, which is a component of a state-of-art ASR system. Meanwhile, the speech attribute pattern is knowledge-driven information integrated with acoustic and phonetic knowledge [25,26]. Here, a set of single activation vector label set are made to describe speech attribute information of each phone in Mandarin. The basic patterns of speech attribute are derived from [27]. We split 2 and 3 vowel transition (2 alias diphthong) into the individual vowels (e.g. iang to i-a-ng) to coarsely simulate the articulatory motions of phoneme production.

The contrastive loss objective in [16] is easy to optimize with a satisfactory performance in the multi-view setup, as listed in Eq. (4). Acoustic feature x and multi-source information y are embedded by network f and g respectively. Fig. 3 shows the embedding model's structure which is the same for network f and g.

$$\min_{f,g} obj^0 := \frac{1}{N}\sum_i^N \max(0, m + d_{cos}(f(x_i^+), g(y_i^+)) - d_{cos}(f(x_i^+), g(y_i^-))) \quad (4)$$

where $f(x)=[f_1(x) f_2(x)]$, $g(y)=[g_1(y) g_2(y)]$, the above objective aims to make the distance between embeddings of paired acoustic feature $x^+$ and information sequence $y^+$ smaller than the distance between embeddings of $x^+$ and an unmatched information sequence $y^-$. Information sequence $y^-$ corresponding to negative phone labels of $x^+$ contrasts with correct information sequence $y^+$. m is a super-parameter representing a margin. $d_{cos}$, as the cosine distance in Eq. (3). Another objective is listed in Eq. (5), where $x^-$ is an unmatched acoustic feature with acoustic feature $x^+$.

$$\min_{f,g} obj^1 := \frac{1}{N}\sum_i^N \max(0, m + d_{cos}(f(x_i^+), g(y_i^+)) - d_{cos}(f(x_i^-), g(y_i^+))) \quad (5)$$

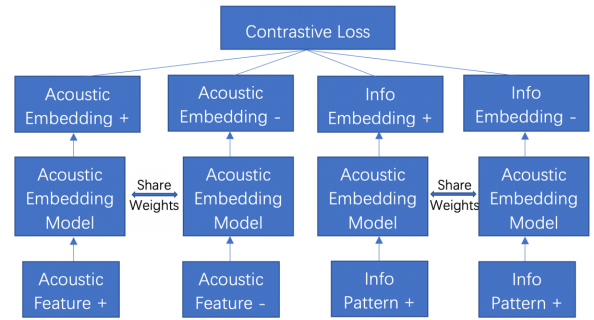

Figure. 2: *multi-view training framework*

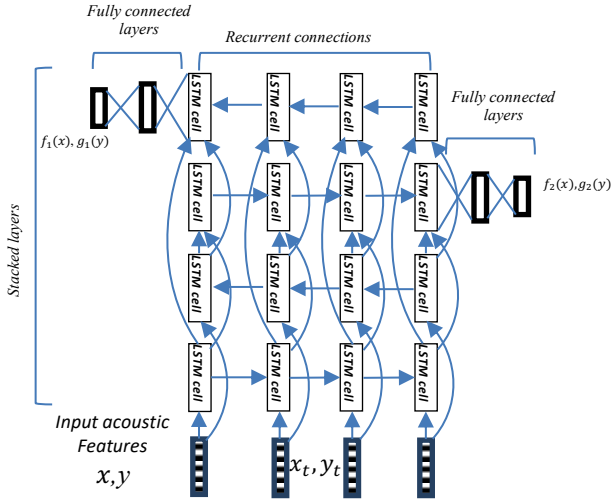

Figure. 3: *embedding model structure*

## 3. Experiments

### 3.1. Speech corpora

The training data employed is from Chinese National Hi-Tech Project 863 [28] for Mandarin large vocabulary continuous speech recognition (LVCSR) system development. Development and test data are from a Chinese L2 speech database, which can be referred to as BLCU inter-Chinese speech corpus [29]. L1 speech data is used as development data, and non-native (L2) speech data is used as test data here.

Table 1: *Training data description*

| Items | information |
|---|---|
| Hours | ≈110h |
| Speaker | 83 L1 males, 83 L1 females |
| Number of utterance | 96745 |
| Number of phonemes | 2351095 |
| Average length per utterance | 12 syllable |

Table 2: dev and *test data description*

| Items | information |
|---|---|
| Hours | ≈13h |
| Speaker | 7 L2 females, 6 L1 males, 6 L1 females |
| Number of utterance | 5469 |
| Number of phonemes | 81142 |
| Average length per utterance | 14 syllable |

### 3.2. GOP-based system setup

We developed a GOP-based assessment system using KALDI [30] to train acoustic models based on an ASR framework of CD–TDNN–HMM, which is also used for forced-alignment and extracting bottleneck features from the sixth TDNN layer. We used 13-dimensional Mel Frequency Cepstrum Coefficient (MFCC) as acoustic features. The TDNN network consists of six hidden layers, each of which contains 850 units. The softmax function is applied to the last layer to produce 2984 (the number of senones) targets of probability distribution function (p.d.f) class. GOP-based system took augmented frame-level feature vectors as input, each of which was composed of 5 preceding, current and 5 succeeding frames, to produce frame-level log-posteriors. When the forced-alignment results were given, Eq. (1) was applied to calculate the GOP scores at phone level, and a threshold of 0.1 was set to tell whether candidate phone was pronounced the way the canonical phone sounded to get the best verification performance.

### 3.3. Single-view system setup

The input to the phone embedding models are acoustic features of 13 dimensions and fixed-length duration of 58 frames, with each phone segment was padded to 58 frames. Before acoustic feature matrixes were put into the models, Cepstrum Mean Variance normalization (CMVN) [31] was applied globally as feature normalization to alleviate influence from the speaker variance. The training set contained about 2.1M example segments, approximate 232k and 153k example segments constituted development and test sets respectively. The training triplets consisted of pairs with same phone types in training set and randomly drawn a third example of each triplets corresponding to a different phone type, as required for the contrastive loss with a margin of 0.4. The network framework is depicted in Fig.1 with three identical embedding models. Bidirectional LSTM was adopted as the embedding model for that it was a natural model class of acoustic phone embedding, since it could handle arbitrary-length sequence feature and each unit of it contained the context-dependent information. Each embedding model consists of two Bidirectional LSTMs. The recurrent hidden state dimension per direction per layer was fixed at 512 and dropout probability [32] of 0.4 was used between stacked recurrent layers. The dimensionalities of the fully connected hidden layers were fixed at 512 and 256 respectively. Dropout probability of 0.4 and Rectified linear unit (RELU) were employed between fully-connected layers. Outputs of the last fully-connected layers were used as the learned embeddings. The training process used the Ada-delta [33] optimizer with an initial learning rate of 0.0001. The best-performing converged model on native speech data was used for mispronunciation verification on the non-native test set.

### 3.4. Multi-view system setup

Embedding models in multi-view network structures for each view are consistent with single-view's setup. For the acoustic view, the inputs of the models were 58*13 matrixes (58 frames for each phone, 13-dimension MFCC feature). The raw bottleneck features were 850 dimensions, then features were processed by dimension reduction into 40-dimensional features at each frame based on Probabilistic Local Pairwise Linear (PLDA) [34]. 58*40 matrixes were taken as the inputs of the data-based view. As shown in [27], there were 31 attribute items to discriminate phones by its speech attribute, for triphthongs and diphthongs, they should be described by 3 individual monophthongs and 2 monophthongs respectively. Therefore, each phones' speech attribute pattern matrix was fixed to 3*31 (consonants, monophthongs and diphthongs were padded with zero from behind). A negative speech attribute label sequence was generated by uniformly sampling a label different from the positive label in training set. Meanwhile, negative acoustic and bottleneck feature sequences were uniformly sampled from all of the mismatched feature sequences in the training set. Each were trained up to 1000 epochs, and AP was computed per 20 epochs.

### 3.5. Evaluation metrics

- False Rejection Rate (FRR): The percentage of mispronunciations being taken as correct pronunciations.
- False Acceptance Rate (FAR): The percentage of correct pronunciations being taken as mispronunciations.
- Diagnostic Accuracy (DA): The percentage of predicted phones correctly recognized i.e. correctly pronounced phones matches correct types, and mispronounced phones were different from the correct ones.
- Averaged precision (AP): By sweeping thresholds, the averaged precision was calculated based on the area under the precision-recall curve.

## 4. Results and Discussion

Two contrastive losses and the combination were adopted to perform the multi-view method. These different objectives were applied to the phone discrimination task on native speech data, Table 3 shows the development set AP on native speech data, using different objectives. $obj^1$ (see Eq.5) slightly outperforms $obj^0$ (see Eq. 4), especially in the method with multi-view of acoustic and bottleneck views. The symmetrized loss function adopts a more comprehensive phonetic information, because of this, the combination of $obj^0$ and $obj^1$ with a symmetrized structure achieves the highest AP, which prominently outperforms the two individual objectives. The embedding-based measures offered acoustic templates at the phone-level, which replaced the likelihood in traditional methods with distance to improve clustering performance. In addition, multi-view methods use multi-source information, which is more discriminative than raw acoustic features, to revise phone-level clustering. The multi-view methods with objective $obj^0 + obj^1$ outperformed the GOP-based method and single-view method, and the single-view method fell short of the multi-view methods, which means the multi-view method made progress on phone-level clustering results over single-view methods. Specifically, the multi-view method with acoustic & speech attributes view achieved the best performance, and the multi-view method with acoustic & bottleneck views was also competitive. Figure 4 & 5 shows the AP on the development set for the multi-view method of acoustic & bottleneck view and acoustic & speech attribute view respectively, using different objectives. As observed, the development set AP grows at a relatively slow rate even after 1000 epochs, and this unsaturated AP indicates that a phone discrimination accuracy could be promoted in a further step. Then the corresponding converged models with the best-performing objective and the fixed threshold of 0.4 were used for mispronunciation verification task on non-native speech data. DAs for various methods were shown in Table 4.

Table 3: *Phone discrimination task with various methods*

| Method | AP | | |
|---|---|---|---|
| GOP | 69.32% | | |
| Single-view | 73.45% | | |
| Multi-view (Acoustic + bottleneck) | $obj^0 + obj^1$ | $obj^0$ | $obj^1$ |
| | 77.16% | 72.46% | 73.82% |
| Multi-view (Acoustic + speech attribute) | $obj^0 + obj^1$ | $obj^0$ | $obj^1$ |
| | 79.41% | 74.33% | 74.61% |

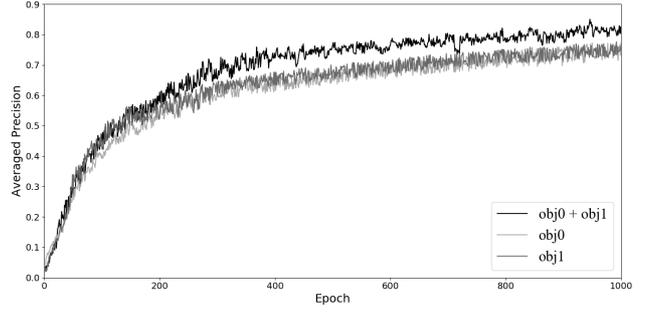

Figure. 4: *dev set AP for different objectives on phone discrimination task (Acoustic + bottleneck)*

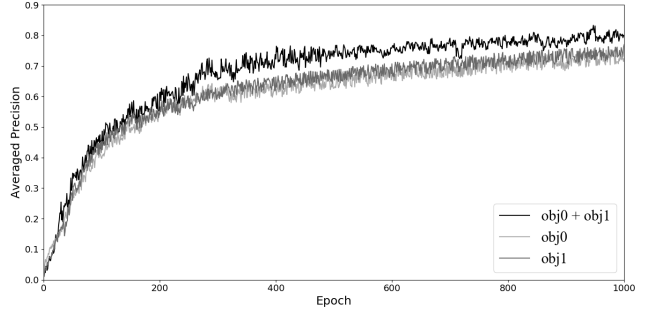

Figure. 5: *dev set AP for different objectives on phone discrimination task (Acoustic + speech attribute)*

Table 4: *mispronunciation verification with various methods*

| Method | FRR | FAR | DA |
|---|---|---|---|
| GOP | 21.86% | 31.36% | 80.93% |
| Single-view | 5.3% | 30.29% | 90.69% |
| Multi-view (Acoustic+bottleneck $obj^0 + obj^1$) | 5.2% | 24.07% | 91.43% |
| Multi-view (Acoustic+speech attribute $obj^0 + obj^1$) | 4.9% | 19.81% | 92.16% |

## 5. Conclusion

In this study, a multi-view approach was considered for a mispronunciation verification task. Acoustic phone embeddings and multi-source information embeddings were jointly learned in the training process, where we had used bottleneck features and speech attribute patterns as multi-source information input views respectively. A range of objectives were explored. GOP-based method and single-view method were considered for comparison. In the single-view method, only the raw acoustic features with the pair-wise labels were used as inputs, it helped to reduce the phone-level recognition errors due to generic acoustic modelling confusion, which means to drop the FRR as depicted in Table 4. While there is still a need for more indicative features to revise the clustering results such that unknown examples are easier to be taken as dissimilar ones with the acoustic templates for a lower FAR. Overall, our final multi-view model of acoustic and speech attribute with combined $obj^0 + obj^1$ shows the best performance over all other approaches.

This work is supported by National social Science foundation of China (18BYY124), Wutong Innovation Platform of Beijing Language and Culture University (16PT05). The corresponding author of the paper is Yanlu Xie.

# 6. References

[1] S. M. Witt and S. J. Young, "Phone-level pronunciation scoring and assessment for interactive language learning", Speech Communication, vol. 30, no. 2-3, pp. 95–108, 2000.
[2] W. Hu, Y. Qian, F. K. Soong, and Y. Wang, "Improved Mispronunciation Detection with Deep Neural Network Trained Acoustic Models and Transfer Learning based Logistic Regression Classifiers", Speech Communication, 67, pp. 154- 166, 2015.
[3] Ellis, R. The Study of Second Language Acquisition. Oxford University Press. 1994.
[4] S. Yoon, M. Hasegawa-Johnson, and R. Sproat, "Landmark Based Automated Pronunciation Error Detection", in Proc. Interspeech, 2010.
[5] R. Duan, et al, "A Preliminary Study on ASR-based Detection of Chinese Mispronunciation by Japanese Learners", in Proc. Interspeech, 2014
[6] Park, Alex S., and J. R. Glass. "Unsupervised Pattern Discovery in Speech", IEEE Transactions on Audio Speech & Language Processing 16.1(2008):186-197.
[7] Jansen, Aren, K. Church, and H. Hermansky. "Towards spoken term discovery at scale with zero resources", in Proc. Interspeech, Makuhari, Chiba, Japan, September DBLP, 2010:1676-1679.
[8] Samy Bengio and Georg Heigold. "Word embeddings for speech recognition", In IEEE Int. Conf. Acoustics, Speech and Sig. Proc., 2014.
[9] Guoguo Chen, Carolina Parada, and Tara N Sainath, "Query-by-example keyword spotting using long short term memory networks", in IEEE International Conference on Acoustics, Speech and Signal Processing (ICASSP), 2015.
[10] Kartik Audhkhasi, Andrew Rosenberg, Abhinav Sethy, Bhuvana Ramabhadran, and Brian Kingsbury. "End-to-end ASR-free keyword search from speech", arXiv preprint arXiv:1701.04313, 2017.
[11] Yu-An Chung, Chao-Chung Wu, Chia-Hao Shen, and Hung-Yi Lee, "Unsupervised learning of audio segment representations using sequence-to-sequence recurrent neural networks", in Proc. Interspeech, 2016.
[12] Herman Kamper, Weiran Wang, and Karen Livescu, "Deep convolutional acoustic word embeddings using word-pair side information", in IEEE International Conference on Acoustics, Speech and Signal Processing (ICASSP), 2016, pp. 4950–4954.
[13] Keith Levin, Katharine Henry, Aren Jansen, and Karen Livescu, "Fixed-dimensional acoustic embeddings of variable-length segments in low-resource settings", in IEEE Automatic Speech Recognition & Understanding (ASRU), 2013.
[14] He, Wanjia , W. Wang , and K. Livescu . "Multi-view Recurrent Neural Acoustic Word Embeddings", arXiv preprint arXiv:1611.04496 (2016).
[15] Settle, Shane, and K. Livescu. " [IEEE 2016 IEEE Spoken Language Technology Workshop (SLT) - San Diego, CA, USA (2016.12.13-2016.12.16)] 2016 IEEE Spoken Language Technology Workshop (SLT) - Discriminative acoustic word embeddings: Recurrent neural network-based approaches", (2016):503-510.
[16] Karl Moritz Hermann and Phil Blunsom. "Multilingual distributed representations without word alignment", In Int. Conf. Learning Representations, 2014. arXiv:1312.6173 [cs.CL].
[17] A. S. Park and J. R. Glass, "Unsupervised pattern discovery in speech", IEEE Trans. Audio, Speech, Language Process., vol.16, no. 1, pp. 186–197, 2008.
[18] A. Jansen and B. Van Durme, "Efficient spoken term discovery using randomized algorithms", in Proc. ASRU, 2011.
[19] J. Bromley, J. W. Bentz, L. Bottou, I. Guyon, Y. LeCun, C. Moore, E. S¨ackinger, and R. Shah, "Signature verification using a "siamese" time delay neural network", International Journal of Pattern Recognition and Artificial Intelligence, vol. 7, no. 04, pp. 669–688, 1993.
[20] R. Hadsell, S. Chopra, and Y. LeCun, "Dimensionality reduction by learning an invariant mapping", in Proc. CVPR, 2006.
[21] P.-S. Huang, X. He, J. Gao, L. Deng, A. Acero, and L. Heck,"Learning deep structured semantic models for web search using clickthrough data", in Proc. CIMK, 2013.
[22] Mikolov T, Chen K, Corrado G, et al. "Efficient Estimation of Word Representations in Vector Space" [J]. Computer Science, 2013.
[23] J. Wieting, M. Bansal, K. Gimpel, and K. Livescu, "From paraphrase database to compositional paraphrase model and back", Trans. ACL, vol. 3, pp. 345–358, 2015.
[24] J. Wieting, M. Bansal, K. Gimpel, and K. Livescu, "From paraphrase database to compositional paraphrase model and back", Trans. ACL, vol. 3, pp. 345–358, 2015.
[25] C.-H. Lee, "From knowledge-ignorant to knowledge-rich modeling: A new speech research paradigm for next generation automatic speech recognition", in Proc. Interspeech, 2004, Jeju Island, Korea, Oct. 4–8, 2004.
[26] C.-H. Lee et al, "An overview on automatic speech attribute transcription (ASAT)", in Proc. Interspeech, 2007, Antwerp, Belgium, Aug. 27–31, 2007, pp. 1825–1828.
[27] Zhang, Chao, Y. Liu, and C. H. Lee. "Detection-based accented speech recognition using articulatory features", *Automatic Speech Recognition & Understanding* IEEE, 2011.
[28] Xu, Bo, et al. "Update Progress of Sinohear: Advanced Mandarin LVCSR System at NLPR", Sixth International Conference on Spoken Language Processing. 2000.
[29] W. Cao, D. Wang J. Zhang, and Z. Xiong, "Developing a Chinese L2 speech database of Japanese learners with narrow phonetic labels for computer assisted pronunciation training", in Proc. Interspeech, 2010.
[30] Povey, Daniel, et al. "The Kaldi speech recognition toolkit", Idiap (2012).
[31] Dahl, George E, et al. "Context-Dependent Pre-Trained Deep Neural Networks for Large-Vocabulary Speech Recognition", IEEE Transactions on Audio Speech & Language Processing 20.1(2012):30-42.
[32] Nitish Srivastava, Geoffrey E Hinton, Alex Krizhevsky, Ilya Sutskever, and Ruslan Salakhutdinov, "Dropout: a simple way to prevent neural networks from overfitting", Journal of Machine Learning Research, vol. 15, no. 1, pp. 1929–1958, 2014.
[33] Matthew D. Zeiler, "ADADELTA: an adaptive learning rate method", CoRR, vol. abs/1212.5701, 2012
[34] Prince, Simon J. D., and J. H. Elder. "Probabilistic Linear Discriminant Analysis for Inferences About Identity", 2007 IEEE 11th International Conference on Computer Vision. IEEE, 2007.